\documentclass[journal]{IEEEtran}

\ifCLASSINFOpdf
\else
   \usepackage[dvips]{graphicx}
\fi
\usepackage{url}

\usepackage{graphicx}
\usepackage{amsmath,amssymb,enumitem,hyperref,booktabs}
\usepackage[numbers]{natbib}

\usepackage{xcolor}
\usepackage[normalem]{ulem}
\usepackage{comment}
\usepackage{multirow}
\usepackage{tikz}
\usetikzlibrary{arrows.meta,positioning,fit,backgrounds}

\hypersetup{hidelinks}

\begin{document}

\title{Do Audio Representations Compose Additively?}

\author{Chenhao Xue,~\IEEEmembership{Student Member,~IEEE,}
Zhijin Guo,
Joyraj Chakraborty,~\IEEEmembership{Member,~IEEE,}
Martin Reed,
and~Nikolaos Thomos,~\IEEEmembership{Senior Member,~IEEE}

\thanks{C. Xue, Z. Guo, and J. Chakraborty are with the Department of Engineering Science, University of Oxford, Oxford, UK (e-mail: \{chenhao.xue, zhijin.guo, joyraj.chakraborty\}@eng.ox.ac.uk). M. Reed and N. Thomos are with the University of Essex, Colchester, UK (e-mail: \{mjreed, nthomos\}@essex.ac.uk)}}

\markboth{}
{Xue \MakeLowercase{\textit{et al.}}: Do Audio Representations Compose Additively?}
\maketitle

\begin{abstract}
Compositionality, the ability to represent complex acoustic scenes as combinations of simpler sound sources, is central to auditory perception and classical additive signal models. Still, it remains unclear whether modern pre-trained audio representations internalize additive structure without compositional supervision. Existing evaluation frameworks of audio compositional reasoning largely focus on cross-modal audio-text alignment, leaving open whether audio representations themselves exhibit additive compositional structure independent of text grounding, analogous to vector arithmetic in word representations. To investigate this, we adopt a two-step diagnostic for frozen audio representations. First, we quantify linear alignment between representations and sound source labels using canonical correlation analysis. Second, we test additive compositional generalization via leave-one-combination-out reconstruction, grouping clips by exact source-label set, averaging their representations, and predicting held-out means from per-source contributions fitted only on training combinations. With larger combination holdouts, CLAP outperforms the permuted and label-overlap baselines on FSD50K, while the speech models do not outperform the label-overlap baseline. We examine representations generated by Wav2Vec2, HuBERT, and CLAP on FSD50K and CHiME-Home datasets. All three models show consistently higher linear correlation and more accurate leave-one-combination-out reconstructions than the permuted baselines. However, only CLAP shows large cosine similarity gains, which could be associated with its training on many kinds of audio and text. Finally, we note that all three models exhibit reconstruction residuals, revealing limits of additive compositionality such as nonlinear or non-compositional audio structure.
\end{abstract}

\begin{IEEEkeywords}
Audio representation learning, additive compositionality, canonical correlation analysis, frozen audio encoders
\end{IEEEkeywords}

\IEEEpeerreviewmaketitle

\vspace{-2mm}\section{Introduction}
\label{sec:intro}

Compositionality, the ability to represent complex structure through combinations of simpler components, is a central principle in intelligent systems~\citep{elmoznino2024complexity}. Human auditory perception similarly organizes complex acoustic scenes into constituent sources, temporal patterns, and semantic attributes~\citep{bregman1994auditory}. Classical audio processing has explicitly exploited such structure through compositional models including non-negative matrix factorization and sparse coding, where signals are represented as combinations of simpler components~\citep{virtanen2015compositional,ozerov2009multichannel,gemmeke2011exemplar,plumbley2009sparse}.

Deep learning has transformed audio processing, with pre-trained representations achieving strong performance across many tasks~\citep{purwins2019deep,liu2022audio,huang2022investigating}. Models such as Wav2Vec2~\citep{baevski2020wav2vec}, HuBERT~\citep{hsu2021hubert}, and CLAP~\citep{elizalde2023clap} learn from self-supervised or audio--text objectives without explicit supervision of how constituent sounds should compose. Their learned representations can nevertheless be difficult to interpret~\citep{rasheed2022explainable,gimeno2025unveiling}, leaving open whether multi-source structure is explicitly reflected in the representation space or arises mainly from statistical associations.

Related questions have been studied extensively in natural language processing. Early work demonstrated linear arithmetic in word~\citep{mikolov2013efficient} and sentence representations~\citep{hewitt2019structural,guo2025quantifying}, motivating additive accounts of semantic composition. However, canonical examples such as $king-man+woman\approx queen$ do not systematically generalize~\citep{church2017word2vec}; non-compositional behavior also appears in analogy and multiword-expression settings~\citep{yazdani2015learning}, motivating nonlinear accounts of composition~\citep{elmoznino2024complexity,wang2024composition}. Related limitations occur in vision-language models, where strong benchmark performance does not necessarily imply correct attribute--object or relational composition~\citep{yuksekgonul2023when}.

Audio compositionality has received increasing attention through source-disentangled representation learning~\citep{sridhar2025compositional} and compositional speaker modeling~\citep{li2021compositional}, but systematic evaluation of additive structure in general-purpose audio representations
remains limited. CompA~\citep{ghosh2023compa} evaluates compositional audio--text alignment through temporal-order and attribute--object tasks,
leaving open whether the audio representations themselves exhibit additive structure independently of text grounding during evaluation. Closely related work by Chen et al.~\citep{chen2026evaluating} evaluates additive structure using synthetic sound scenes. We instead study real recordings with semantic sound-source labels and ask whether representations of unseen source combinations can be reconstructed from source contributions estimated only from other combinations. Additive structure also has practical value: if a mixture representation can be approximated from source-specific contributions,
an unseen combination of known sources can be estimated in representation space without recording that exact mixture or fine-tuning the encoder.

\noindent Our main contributions can be summarized as:
\begin{itemize}[leftmargin=*, nosep]
    \item we formulate additive compositionality in frozen audio representations and evaluate it through linear alignment and reconstruction of unseen source combinations;
    \item we compare Wav2Vec2, HuBERT, and CLAP on real recordings from FSD50K and CHiME-Home using semantic multi-hot source labels; and
    \item we introduce a stronger combination-disjoint repeated-holdout evaluation with permutation and label-overlap baselines, revealing when reconstruction success does not extend beyond source-label overlap.
\end{itemize}

\vspace{-3mm}\section{Problem Formulation}
\label{sec:prob_form}


In this paper, we address the question: \emph{Do pre-trained audio representation models encode compositional structure where complex acoustic scenes decompose as additive combinations of constituent sound sources?}


We consider a collection of audio clips $\mathcal{C} = \{c_1,\allowbreak c_2,\allowbreak \ldots,\allowbreak c_M\}$ from real-world acoustic scenes, where $M$ is the number of clips. Clips may contain co-occurring sound sources (e.g., human speech, bird chirps, traffic noise, music). Let $\mathcal{S} = \{s_1,\allowbreak s_2,\allowbreak \ldots,\allowbreak s_n\}$ denote the set of all sound source types labeled in the collection, where $n$ is the number of types.

Given a pre-trained, frozen audio encoder $f: \mathcal{C} \to \mathbb{R}^m$ that maps each clip to an $m$-dimensional representation, we investigate whether the representation space exhibits \emph{additive compositional structure}. For reconstruction, clips with identical source labels are grouped and their representations averaged, giving one row per unique source combination. Holding out a row therefore tests a combination absent from the fitting data. Let $q$ denote the number of unique combinations and $S \in \{0,1\}^{q \times n}$ a multi-hot binary matrix, where $S_{ij}=1$ if combination $i$ contains source $s_j$, and 0 otherwise. We further denote $R \in \mathbb{R}^{q \times m}$ as the representation matrix, where row $r_i$ is the mean representation of the clips in combination $i$. The additive compositionality hypothesis posits that there exists a source contribution weight matrix $W \in \mathbb{R}^{n \times m}$ such that:
\begin{equation}
\label{eq:problem_additive}
SW \approx R,
\end{equation}
where each row $w_j$ is the fitted contribution of source $s_j$. Hence, if combination $i$ contains sources $\{s_j, s_k, s_l\}$, then $r_i \approx w_j + w_k + w_l$.

This formulation is motivated by both physical and computational considerations. From a physical perspective, acoustic signals are superimposed, and hence the waveform of a mixture is the sum of component waveforms. In power spectral representations, uncorrelated sources contribute additively to power. This property has been exploited by classical methods such as non-negative matrix factorization for source separation. From a computational perspective, additive compositionality implies that each source $s_j$ has a context-independent contribution $w_j$. In other words, the contribution of a dog bark to the representation should be similar, independently of whether it co-occurs with traffic noise or bird chirps.

The proposed formulation differs fundamentally from prior work on audio compositionality evaluation. Unlike CompA~\citep{ghosh2023compa}, which evaluates cross-modal compositional alignment (i.e., whether models correctly match audio to compositional text descriptions), we assess compositional structure within the audio representation space itself, independent of text grounding during evaluation. Human labels are used to construct $S$ and fit $W$; the encoders remain frozen. The central question is whether and to what extent pre-trained audio representations encode linear, additive compositional structure aligned with human-interpretable sound source labels.

  \vspace{-2mm} \section{Methodology}
\label{sec:methods}

\begin{figure*}[t]
\centering
\includegraphics[width=0.75\linewidth]{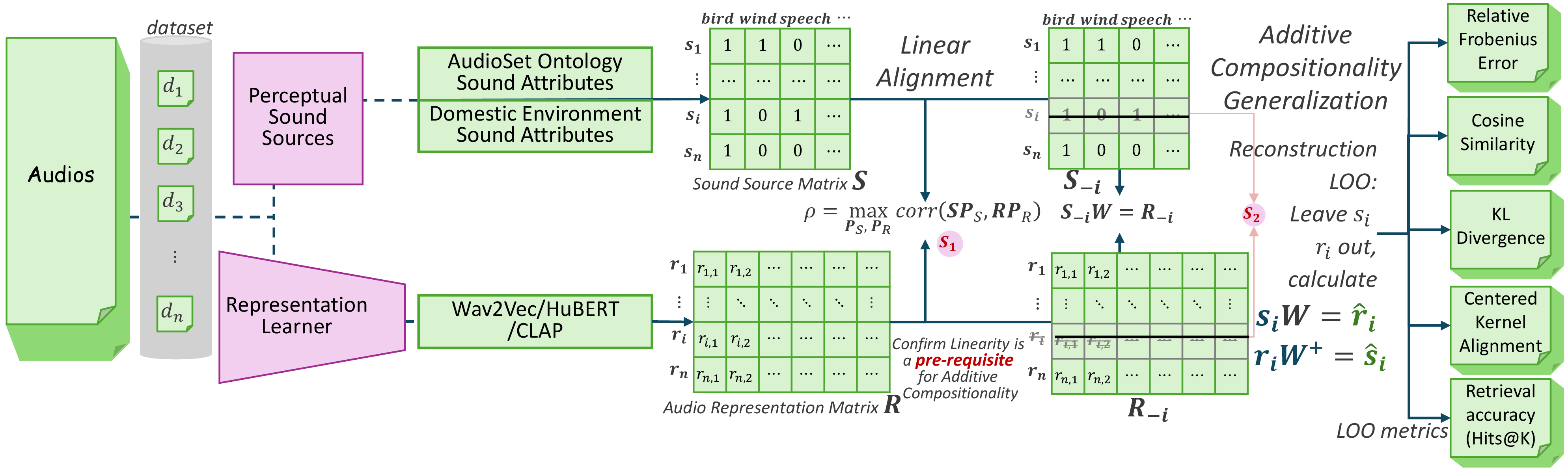}

\caption{Two-step evaluation framework for additive compositionality in audio representations. Step 1 (top): CCA quantifies linear alignment between the source labels and the representations. Step 2 (bottom): Leave-one-combination-out reconstruction tests whether a held-out combination can be predicted from source contributions learned on other combinations.}
\label{fig:method}
\end{figure*}

In our work, we adapt the two-step compositional diagnostic from~\citep{guo2025quantifying, xu2023compositionality} to audio representations. As illustrated in Fig.~\ref{fig:method}, this framework evaluates both linear alignment and additive generalization for compositional structure.

\vspace{-3mm}\subsection{Compositional Diagnostic}

To examine the compositionality, we follow the process outlined below:

\noindent\textbf{Step 1 - Linear alignment:} We quantify the linear correlation between the source labels and the representations using Canonical Correlation Analysis (CCA). For this alignment test, each row is one clip, not a grouped combination. CCA learns projections $P_S$ and $P_R$ that maximize the correlation $\rho = \max_{P_S, P_R} \mathrm{corr}(SP_S, RP_R)$. We, then, report the canonical correlation at each projection dimension.


\noindent\textbf{Step 2 - Additive generalization:} While CCA establishes linear correlation, it does not test whether representations generalize compositionally to new source combinations. We employ a leave-one-combination-out (LOO) protocol. Specifically, for each combination $i$, we estimate the source contribution matrix $W$ on all combinations except $i$, and then we predict the held-out representation as $\hat{\mathbf{r}}_i = \mathbf{s}_i W$. Each held-out row represents a source combination excluded from fitting $W$. This additive reconstruction follows the tree reconstruction error framework of Andreas~\citep{andreas2019measuring}.

\noindent\textbf{Significance testing:} For both steps, we compare against a row-permuted baseline that shuffles $S$'s rows, impairing row-label alignment while preserving label co-occurrence statistics. We use $N=100$ permutations, reporting their mean for CCA and their best score (or minimum loss) for each reconstruction metric. In repeated combination holdouts (Table~\ref{tab:reconstruction}(b)), only training pairings are permuted, and predictions are evaluated against true held-out representations. For a statistic $T$, with $T^{(b)}$ its value under permutation $b$, the Monte-Carlo $p$-value is:
\begin{equation}
p \;=\; \frac{1 + \sum_{b=1}^{N} \mathbb{I}\!\big[T^{(b)} \ge T^{\mathrm{real}}\big]}{N+1},
\end{equation}
with the inequality reversed for loss-type statistics. For repeated holdouts only, overlap predictors either copy the maximum-Jaccard training-combination representation or average training representations with Jaccard weights. Jaccard overlap is the ratio of intersection to union sizes of source-label sets; we report the better predictor per metric.

\vspace{-4mm}\subsection{Ridge Regularization for Hierarchical Audio Labels}\label{sec:ridge}

Audio tags exhibit properties that require adaptation beyond the standard formulation. Sound source labels often follow a hierarchy (e.g., \textit{Animal} $\supset$ \textit{Wild\_animal} $\supset$ \textit{Bird}) and co-occur across semantic families, e.g., \textit{Speech} with \textit{Music} in vocal performances. This hierarchical structure induces strong correlations among columns of $S$, which can cause numerical instability in the standard pseudoinverse solution.

A ridge-regularized pseudoinverse addresses this:
\begin{equation}
\label{eq:ridge}
W_{\lambda} = (S^{\top}S + \lambda I)^{-1}S^{\top}R,
\end{equation}
where $\lambda$ is the regularization parameter. This estimator shrinks unstable directions in correlated subspaces while converging to the Moore--Penrose pseudoinverse as $\lambda \to 0$. Importantly, our diagnostic targets linear structure rather than statistical independence: correlated columns in $S$ do not invalidate either CCA or additive reconstruction. The regularization ensures numerical stability without altering the fundamental hypothesis being tested. CHiME-Home uses centered ridge with a validation-selected, fixed $\lambda$; FSD50K uses the uncentered pseudoinverse. In centered fits, training means are subtracted from labels and representations, and the training representation mean is added back to predictions.

\vspace{-3mm}\subsection{Reconstruction Metrics}
We evaluate representation reconstruction quality using five metrics, each capturing different aspects of compositional fidelity. For LOO, let $\hat{R} \in \mathbb{R}^{q \times m}$ denote the matrix of predictions; repeated-holdout metrics are computed on the held-out rows:

\begin{enumerate}[leftmargin=1.5em,topsep=2pt,itemsep=2pt]
    \item \textbf{Relative Frobenius error:} $\|\hat{R} - R\|_F / \|R\|_F$ measures reconstruction error as a proportion of original signal magnitude, providing scale-invariant assessment.
    \item \textbf{Cosine similarity:} $\langle \hat{\mathbf{r}}_i, \mathbf{r}_i \rangle / (\|\hat{\mathbf{r}}_i\| \|\mathbf{r}_i\|)$, averaged over all combinations, measures directional alignment between reconstructed and true representations.
    \item \textbf{KL divergence:} $D_{\text{KL}}(p_i \| q_i) = \sum_{j=1}^{m} p_{ij} \log_2(p_{ij}/q_{ij})$ where $p_i = \text{softmax}(r_i)$ and $q_i = \text{softmax}(\hat{r}_i)$, quantifying distributional divergence in bits, averaged over all combinations.
    \item \textbf{Centered Kernel Alignment (CKA):} $\mathrm{CKA}(R, \hat{R}) = \mathrm{HSIC}(K, L) / \sqrt{\mathrm{HSIC}(K, K) \cdot \mathrm{HSIC}(L, L)}$ where $K = RR^{\top}$ and $L = \hat{R}\hat{R}^{\top}$ are linear kernels double-centred by HSIC (the Hilbert-Schmidt Independence Criterion), which measures statistical dependence between kernel matrices~\citep{gretton2005measuring,kornblith2019similarity}, assessing whether pairwise similarity structure is preserved.
    \item \textbf{Nearest-neighbor retrieval (Hits@K):} The fraction of cases where the true representation $\mathbf{r}_i$ appears among the top-$K$ nearest neighbors of $\hat{\mathbf{r}}_i$, testing whether the prediction is close enough to retrieve the correct held-out combination. LOO ranks all $q$ combinations by cosine similarity; repeated holdouts rank only the split's held-out combinations by Euclidean distance.
\end{enumerate}

Relative Frobenius error measures normalized reconstruction error; cosine similarity measures direction; CKA compares pairwise similarities; KL divergence compares softmax distributions; and Hits@K measures retrieval accuracy.

\vspace{-2mm}\section{Experiments and Results}
\label{sec:exp}

\subsection{Models, Datasets, and Setups}

\begin{figure*}[t]
\centering
\includegraphics[width=0.92\linewidth]{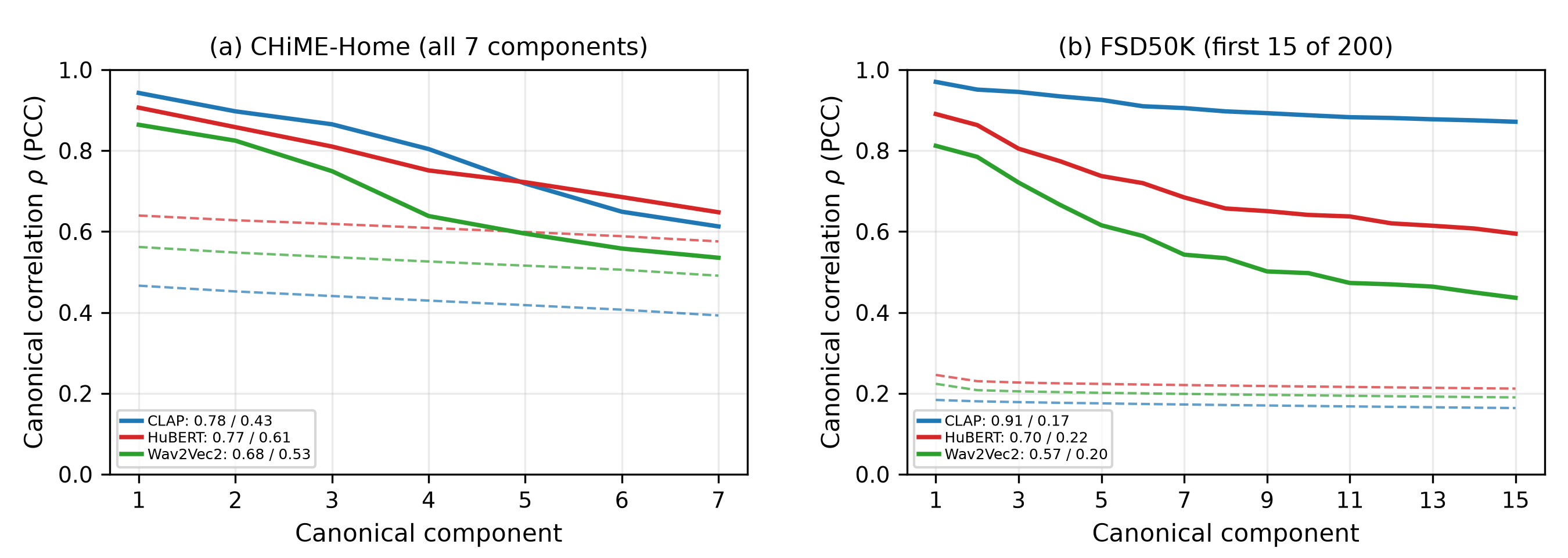}
\caption{Canonical correlations for CHiME-Home (7 components) and FSD50K (first 15 of 200). Solid: real; dashed: mean of 100 permutations. Legends give mean Pearson correlations (PCCs), real / permuted.}
\label{fig:cca}
\end{figure*}

We evaluate representations from Wav2Vec2~\citep{baevski2020wav2vec}, HuBERT~\citep{hsu2021hubert}, and CLAP~\citep{elizalde2023clap}. Wav2Vec2 and HuBERT are self-supervised models trained on speech through contrastive and masked prediction objectives. CLAP learns joint audio-text representations through contrastive learning on diverse audio sources, including environmental, urban, music, and speech sounds. We extract one representation per clip: CLAP uses \nolinkurl{630k-audioset-best.pt} without fusion and provides a 512-dimensional projected audio representation; HuBERT (\nolinkurl{facebook/hubert-large-ls960-ft}, 1024-dimensional) and Wav2Vec2 (\nolinkurl{facebook/wav2vec2-base-960h}, 768-dimensional) use mean-pooled last hidden states. The source contribution matrix is fit only after the representations are extracted. All encoders remain frozen throughout the analysis.

We use FSD50K~\citep{fonseca2022FSD50K} and CHiME-Home~\citep{foster2015chime} datasets for our investigation. FSD50K contains 51,197 audio clips with an average length of 7.6 seconds, labeled with 200 sound classes organized hierarchically from the AudioSet Ontology~\citep{gemmeke2017audio}, covering human sounds, environmental sounds, animal sounds, musical instruments, and mechanical sounds. CHiME-Home provides 2,762 refined 4-second chunks (1,946 development and 816 evaluation), which we analyze using the seven content classes. The short clip duration makes it more likely that labeled sound sources co-occur within single acoustic scenes rather than across temporally distinct events. The LOO analysis uses 39,680 FSD50K clips forming 1,289 combinations and 2,762 CHiME-Home chunks forming 67 combinations.

For ridge regularization, $\lambda$ is selected once from 50 logarithmically spaced values in $[10^{-6}, 10^{8}]$ by minimizing mean squared error on a held-out validation split, then held fixed across LOO folds. Repeated holdouts use fixed per-encoder regularization. For nearest-neighbor retrieval (Hits@K), we set $K=5$.

\vspace{-3mm}\subsection{Quantifying Linear Alignment}

Figure~\ref{fig:cca} shows higher canonical correlations for the true source--representation pairings than for permutation baselines across all three models and both datasets. The separation is larger on FSD50K and is most pronounced for CLAP. These results establish linear source-label alignment, but do not by themselves establish additive generalization.

\begin{table*}[t]
\centering
\caption{Held-out combination reconstruction.
(a) LOO: additive / permutation.
(b) Repeated holdout: additive mean$\pm$SD / permutation / label-overlap
over 20 splits. Bold in (b) indicates additive means better than both baselines
before rounding.}
\label{tab:reconstruction}
\label{tab:loo}
\label{tab:combodisjoint}
\scriptsize\setlength{\tabcolsep}{2pt}\renewcommand{\arraystretch}{0.9}
\setlength{\aboverulesep}{1pt}\setlength{\belowrulesep}{1pt}
\begin{tabular*}{\textwidth}{@{\extracolsep{\fill}}llccccc@{}}
\toprule
Dataset & Model & Cos.~Sim.~($\uparrow$) & Rel.~Frob.~($\downarrow$) & CKA~($\uparrow$) & KL~($\downarrow$) & Hits@5~($\uparrow$) \\
\midrule
\multicolumn{7}{l}{\textbf{(a) Leave-one-combination-out}} \\
\midrule
\multirow{3}{*}{FSD50K} & CLAP & 0.686 / 0.270 & 0.761 / 1.015 & 0.579 / 0.032 & 0.0007 / 0.0012 & 0.43 / 0.01 \\
 & HuBERT & 0.968 / 0.957 & 0.410 / 0.436 & 0.042 / 0.006 & 0.0093 / 0.0105 & 0.06 / 0.01 \\
 & Wav2Vec2 & 0.942 / 0.930 & 0.470 / 0.501 & 0.055 / 0.005 & 0.0065 / 0.0075 & 0.04 / 0.01 \\
\midrule
\multirow{3}{*}{CHiME-Home} & CLAP & 0.899 / 0.820 & 0.455 / 0.586 & 0.714 / 0.196 & 0.0002 / 0.0004 & 0.58 / 0.15 \\
 & HuBERT & 0.992 / 0.988 & 0.134 / 0.159 & 0.415 / 0.269 & 0.0009 / 0.0013 & 0.15 / 0.13 \\
 & Wav2Vec2 & 0.975 / 0.971 & 0.239 / 0.265 & 0.301 / 0.212 & 0.0011 / 0.0014 & 0.21 / 0.15 \\
\midrule
\multicolumn{7}{l}{\textbf{(b) Repeated combination holdout}} \\
\midrule
\multirow{3}{*}{FSD50K} & CLAP & \textbf{0.67{\tiny$\pm$0.01}} / 0.27 / 0.61 & \textbf{0.78{\tiny$\pm$0.01}} / 1.03 / 0.81 & \textbf{0.60{\tiny$\pm$0.02}} / 0.21 / 0.52 & \textbf{0.0007{\tiny$\pm$$<$0.0001}} / 0.0013 / 0.0008 & \textbf{0.57{\tiny$\pm$0.02}} / 0.04 / 0.52 \\
 & HuBERT & 0.96{\tiny$\pm$$<$0.01} / 0.96 / 0.97 & 0.42{\tiny$\pm$0.01} / 0.43 / 0.25 & 0.05{\tiny$\pm$0.01} / 0.01 / 0.17 & 0.0097{\tiny$\pm$0.0006} / 0.0105 / 0.0034 & 0.08{\tiny$\pm$0.02} / 0.04 / 0.14 \\
 & Wav2Vec2 & 0.93{\tiny$\pm$$<$0.01} / 0.93 / 0.95 & 0.48{\tiny$\pm$0.01} / 0.50 / 0.35 & 0.06{\tiny$\pm$0.02} / 0.02 / 0.13 & 0.0068{\tiny$\pm$0.0004} / 0.0073 / 0.0037 & 0.07{\tiny$\pm$0.01} / 0.04 / 0.11 \\
\midrule
\multirow{3}{*}{CHiME-Home} & CLAP & \textbf{0.90{\tiny$\pm$0.02}} / 0.85 / 0.86 & \textbf{0.46{\tiny$\pm$0.04}} / 0.54 / 0.52 & \textbf{0.76{\tiny$\pm$0.06}} / 0.73 / 0.62 & \textbf{0.0002{\tiny$\pm$$<$0.0001}} / 0.0003 / 0.0003 & \textbf{0.82{\tiny$\pm$0.09}} / 0.63 / 0.66 \\
 & HuBERT & 0.99{\tiny$\pm$$<$0.01} / 0.99 / 0.99 & 0.13{\tiny$\pm$0.02} / 0.13 / 0.14 & 0.48{\tiny$\pm$0.10} / 0.59 / 0.46 & 0.0009{\tiny$\pm$0.0002} / 0.0009 / 0.0011 & \textbf{0.65{\tiny$\pm$0.07}} / 0.62 / 0.42 \\
 & Wav2Vec2 & 0.98{\tiny$\pm$0.01} / 0.98 / 0.98 & 0.22{\tiny$\pm$0.04} / 0.22 / 0.23 & 0.41{\tiny$\pm$0.15} / 0.56 / 0.45 & \textbf{0.0010{\tiny$\pm$0.0004}} / 0.0010 / 0.0011 & 0.59{\tiny$\pm$0.07} / 0.59 / 0.45 \\
\bottomrule
\end{tabular*}
\end{table*}

\vspace{-3mm}\subsection{Quantifying Additive Generalization}

Having confirmed linear correlation, we next evaluate whether representations can be additively reconstructed from their constituent sound sources. Table~\ref{tab:reconstruction}(a) reports leave-one-combination-out results for the five metrics in Section~\ref{sec:methods}. All three models outperform the permuted baseline on all five metrics across both datasets; cosine-similarity gains are large only for CLAP.
We hold out approximately 20\% of combinations in 20 random splits, moving combinations back to training when needed to keep all test sources in the training data. Table~\ref{tab:reconstruction}(b) reports means and standard deviations across these splits (CHiME-Home: 2,720 clips and 66 combinations after excluding 42 empty-label clips; FSD50K: 1,289 combinations). On FSD50K, CLAP outperforms both the permuted and label-overlap baselines on all five metrics (CKA 0.60 versus 0.21 and 0.52). The speech models perform worse than label overlap on relative Frobenius error, CKA, and retrieval; their high raw cosine does not establish an additive advantage. CHiME-Home has a smaller test pool and mixed gains across measures: CLAP's CKA is 0.76 compared with 0.73 for the permuted baseline and 0.62 for label overlap, a small mean gain over the permuted baseline. Given the sources, the mean representation of an unseen combination can thus be approximated by summing source contributions. This is more evident for CLAP and is attributed to its audio-text training and differences in architecture and objective.




\vspace{-3mm}\section{Conclusions}


Our work was motivated by the question \emph{`Is an audio model's view of a sound scene close to the sum of its sources?'}. Our investigation found that there is strong evidence for CLAP, while for Wav2Vec2 and HuBERT there is only partial evidence. To address this, in this work, we formulated the task of evaluating additive compositional generalization in audio representations and adapted a two-step diagnostic with CCA and leave-one-combination-out reconstruction. We evaluated the three models under consideration on FSD50K and CHiME-Home. The results showed linear correlations with sound-source labels and better reconstruction than the permuted baseline in LOO. CLAP also outperformed label overlap on FSD50K in repeated combination holdouts. Given the sources, unseen mean representations are thus additively predictable, most clearly for CLAP. Beyond detecting additive compositionality, our framework also quantifies reconstruction residuals. All three models fall short of ideal reconstruction, highlighting limits of linear compositionality. Acoustic effects such as reverberation and masking, together with nonlinear encoder processing, may contribute to these residuals. Semantic structure may encode coherent scenes rather than independent source sums. Our analysis treats each clip as an unordered set of sources and does not model temporal arrangement. On FSD50K clips of at most 4.5 seconds, CLAP also exceeds the permuted baseline (cosine 0.68 versus 0.28).

 \bibliographystyle{IEEEtran}
\bibliography{refs}

\end{document}


\maketitle

\tableofcontents
\newpage

\section{Overview}

The main paper evaluates additive compositionality in audio representations using a two-step diagnostic: (1) CCA for linear alignment, and (2) leave-one-out (LOO) reconstruction for additive generalization. Both steps compare real source-representation pairings against row-permuted baselines.

This supplementary document describes iterative experiments that investigate an important question: \emph{under what conditions does this evaluation framework produce valid results?} Specifically, we ask:
\begin{itemize}[nosep]
    \item How many data instances (clips) are needed for the LOO evaluation to be meaningful?
    \item What happens when data becomes scarce?
    \item Is the evaluation validity a property of the data, the model, or both?
\end{itemize}

We address these questions through two complementary experiments --- iterative removal and iterative build --- applied across five dataset configurations.

\section{Background: The Evaluation Framework}
\label{sec:background}

\subsection{Core formulation}

The additive compositionality hypothesis (Equation~1 in the main paper) states that there exists a source contribution weight matrix $W \in \mathbb{R}^{n \times m}$ such that $SW \approx R$, where $S \in \{0,1\}^{q \times n}$ is the multi-hot source label matrix and $R \in \mathbb{R}^{q \times m}$ is the representation matrix.

We solve for $W$ using a ridge-regularized pseudoinverse:
\begin{equation}
W_{\lambda} = (S^{\top}S + \lambda I)^{-1}S^{\top}R.
\end{equation}

The LOO evaluation holds out one clip $i$, estimates $W$ on the remaining clips, and predicts $\hat{r}_i = s_i W$. If the prediction is close to the true $r_i$, and closer than when $S$ is randomly permuted, the representation exhibits additive compositionality.

\subsection{The compositionality gap}

Given $N = 100$ row-permuted baselines, we define the best-of-random baseline as $T^* = \max_b T^{(b)}$ for quality-type metrics (CKA, cosine similarity, Hits@K) and $T^* = \min_b T^{(b)}$ for error-type metrics (relative Frobenius error, L2 norm, KL divergence).

The compositionality gap is:
\begin{equation}
\Delta_T = \begin{cases}
T^{\text{real}} - T^* & \text{if } T \text{ is quality-type} \\
T^* - T^{\text{real}} & \text{if } T \text{ is error-type}
\end{cases}
\end{equation}

Under both conventions, $\Delta_T > 0$ means the true $(S, R)$ pairing produces better LOO reconstruction than the best of $N$ random pairings. Using best-of-$N$ (not mean-of-$N$) makes the baseline conservative: among 100 permutations, some may achieve good reconstruction by chance, and comparing against the best accounts for this.

\subsection{The validity problem}

The LOO evaluation is a statistical test. Like any statistical test, it has assumptions. The key assumption here is that the linear system $SW \approx R$ is \emph{overdetermined}: the source matrix $S$ must have more rows (clips) than its effective dimensionality. When this assumption is violated --- when $S$ has too few rows relative to its rank --- the pseudoinverse can fit any target near-perfectly, including permuted baselines. In this regime, all metrics become trivially good for both real and random pairings, and the compositionality gap collapses to zero. The evaluation loses its ability to distinguish genuine compositionality from numerical artifacts.

\subsection{Overdetermination ratio}

We quantify the constraint level using the overdetermination ratio:
\begin{equation}
\mathcal{R} = \frac{q}{\text{eff\_rank}(S)},
\end{equation}
where $\text{eff\_rank}(S) = \exp(H(p))$, with $p$ being the normalized singular value distribution of $S$ \cite{roy2007effective}. When $\mathcal{R} \gg 1$, the system is overdetermined and the evaluation is meaningful. When $\mathcal{R}$ approaches 1, the system becomes underdetermined.

\section{Datasets and Configurations}
\label{sec:datasets}

We run iterative experiments on five dataset configurations:

\begin{enumerate}[nosep]
    \item \textbf{FSD50K (full):} 1289 clips, 200 sound classes. The same dataset used in the main paper.
    \item \textbf{FSD50K Environmental:} 758 clips from the Environmental sound category of FSD50K.
    \item \textbf{FSD50K Music:} 81 clips from the Music category of FSD50K.
    \item \textbf{FSD50K Speech:} 89 clips from the Speech category of FSD50K.
    \item \textbf{CHiME-Home:} 67 clips with 7 sound classes. An independent dataset.
\end{enumerate}

\begin{table}[h]
\centering
\caption{Dataset properties at full size.}
\label{tab:dataset_properties}
\begin{tabular}{lrrrrr}
\toprule
Config & $q$ & $n$ & eff\_rank($S$) & $\mathcal{R}_{\text{start}}$ \\
\midrule
FSD50K       & 1289 & 200 & $\sim$150 & $\sim$8.6 \\
FSD50K Env   & 758  & 127  & $\sim$97  & $\sim$7.8 \\
FSD50K Music & 81   & 36   & $\sim$25  & $\sim$3.3 \\
FSD50K Speech& 89   & 27   & $\sim$21  & $\sim$4.2 \\
CHiME-Home   & 67   & 9   & $\sim$7   & $\sim$9.8 \\
\bottomrule
\end{tabular}
\end{table}

All configurations start with $\mathcal{R} > 3$, meaning the LOO evaluation is valid at full dataset size. The iterative experiments reveal at what point this validity breaks down.

The three models evaluated are the same as in the main paper: CLAP, HuBERT, and Wav2Vec2.

\section{Iterative Removal Experiment}
\label{sec:removal}

\subsection{Design}

The iterative removal experiment starts from the full dataset ($q$ clips) and removes one row of $S$ (and the corresponding row of $R$) at each iteration. At each step, the full LOO evaluation is recomputed on the reduced dataset. This reveals how the evaluation degrades as data becomes scarce.

Three removal orderings are compared:

\begin{itemize}[nosep]
    \item \textbf{Entropy ordering:} removes the row with the smallest entropy contribution to $S$ first. This ordering uses only $S$ and is therefore the same for all three models.
    \item \textbf{Effective-columns (effcols) ordering:} removes rows by their contribution to the number of active columns in $S$. Also uses only $S$.
    \item \textbf{OMP ordering:} removes the row whose representation is best predicted by the current linear model (i.e., the row with the smallest reconstruction residual in the $S \to R$ mapping). This ordering uses both $S$ and $R$, so it differs across models. OMP stands for Orthogonal Matching Pursuit, reflecting its greedy residual-based selection.
\end{itemize}

\subsection{Three thresholds identified}

For each experiment, we identify three thresholds by inspecting per-iteration metric values:

\begin{enumerate}[nosep]
    \item \textbf{values\_diminish:} the number of remaining rows below which metric values cease to be numerically meaningful. Below this point, metrics are dominated by numerical artifacts.
    \item \textbf{turning\_point:} the number of remaining rows at which cosine similarity between real and reconstructed representations reaches its lowest point (if the trajectory is non-monotonic). If cosine similarity decreases monotonically, there is no turning point.
    \item \textbf{gap\_positive\_until:} the number of remaining rows down to which the compositionality gap remains positive. Below this point, the real pairing no longer outperforms the best random permutation.
\end{enumerate}

These thresholds form a natural ordering: gap\_positive\_until $>$ turning\_point $>$ values\_diminish (reading in the direction of decreasing rows). This means the gap becomes negative before the turning point is reached, and the turning point occurs before metrics become meaningless.

\subsection{Results}

\subsubsection{Entropy and effcols produce identical turning points}

For both entropy and effcols orderings, the turning point is the same for all three models within each config. This is expected because these orderings depend only on $S$, which is shared across models.

\begin{table}[h]
\centering
\caption{Entropy/effcols removal: values\_diminish and turning\_point (rows remaining). These are model-independent within each config.}
\label{tab:removal_entropy}
\begin{tabular}{lrr}
\toprule
Config & values\_diminish & turning\_point \\
\midrule
FSD50K       & 139 & 140 \\
FSD50K Env   & 87  & 90--92 \\
FSD50K Music & 24  & 30--34 \\
FSD50K Speech& 25  & 23--28 \\
CHiME-Home   & 5--6 & 6--8 \\
\bottomrule
\end{tabular}
\end{table}

\subsubsection{OMP delays the turning point}

OMP removal reaches the turning point at far fewer remaining rows than entropy/effcols. This is because OMP preferentially removes redundant rows (those well-explained by the current model), which preserves linear independence and keeps the condition number low.

\begin{table}[h]
\centering
\caption{OMP removal: values\_diminish and turning\_point (rows remaining) by model. ``3'' indicates values remain meaningful down to the experiment minimum.}
\label{tab:removal_omp}
\begin{tabular}{l rr rr rr}
\toprule
 & \multicolumn{2}{c}{CLAP} & \multicolumn{2}{c}{HuBERT} & \multicolumn{2}{c}{Wav2Vec2} \\
\cmidrule(lr){2-3} \cmidrule(lr){4-5} \cmidrule(lr){6-7}
Config & dim. & t.p. & dim. & t.p. & dim. & t.p. \\
\midrule
FSD50K       & 3    & None & 20   & 89  & 17   & 119 \\
FSD50K Env   & 3    & None & 68   & 69  & 48   & 75  \\
FSD50K Music & 3    & None & 5    & 18  & 3    & 15  \\
FSD50K Speech& 3    & None & 9    & 10  & 3    & 14  \\
CHiME-Home   & 3    & None & 4    & 4   & 5    & 8   \\
\bottomrule
\end{tabular}
\end{table}

\subsubsection{CLAP OMP shows no turning point in any config}

Across all five configurations, CLAP OMP removal produces monotonic cosine similarity: the metric never dips, and values remain meaningful until the experiment minimum (3 rows). This means CLAP's compositional structure is so strong that even with very few carefully selected rows, the linear model captures the correct direction in representation space.

No other model-method combination shows this universal monotonicity.

\subsubsection{Gap-positive threshold reveals model differences}

While entropy/effcols turning points are model-independent (governed by $S$ alone), the gap\_positive\_until threshold varies across models:

\begin{table}[h]
\centering
\caption{Entropy removal: gap\_positive\_until (rows remaining) by model.}
\label{tab:removal_gap}
\begin{tabular}{lrrr}
\toprule
Config & CLAP & HuBERT & Wav2Vec2 \\
\midrule
FSD50K       & 141  & 270  & 450 \\
FSD50K Env   & 88   & 150  & 184 \\
FSD50K Music & 34   & 65   & 72 \\
FSD50K Speech& 28   & 47   & 47 \\
CHiME-Home   & 21   & 43   & 47 \\
\bottomrule
\end{tabular}
\end{table}

CLAP maintains a positive gap down to fewer remaining rows than HuBERT or Wav2Vec2 in every config. This reflects CLAP's stronger compositional signal, which is harder for random permutations to match.

The effcols ordering produces nearly identical gap thresholds:

\begin{table}[h]
\centering
\caption{Effcols removal: gap\_positive\_until (rows remaining) by model.}
\label{tab:removal_gap_effcols}
\begin{tabular}{lrrr}
\toprule
Config & CLAP & HuBERT & Wav2Vec2 \\
\midrule
FSD50K       & 141  & 260  & 450 \\
FSD50K Env   & 88   & 150  & 184 \\
FSD50K Music & 34   & 65   & 72 \\
FSD50K Speech& 31   & 47   & 47 \\
CHiME-Home   & 21   & 43   & 47 \\
\bottomrule
\end{tabular}
\end{table}

OMP removal preserves the gap much longer, because it removes redundant rows first:

\begin{table}[h]
\centering
\caption{OMP removal: gap\_positive\_until (rows remaining) by model.}
\label{tab:removal_gap_omp}
\begin{tabular}{lrrr}
\toprule
Config & CLAP & HuBERT & Wav2Vec2 \\
\midrule
FSD50K       & 39   & 750  & 750 \\
FSD50K Env   & 54   & 400  & 400 \\
FSD50K Music & 36   & 69   & 79 \\
FSD50K Speech& 25   & 69   & 62 \\
CHiME-Home   & 19   & 42   & 42 \\
\bottomrule
\end{tabular}
\end{table}

For FSD50K and FSD50K Environmental, OMP delays the gap collapse dramatically: CLAP OMP maintains a positive gap down to 39 and 54 rows remaining respectively, compared to 141 and 88 for entropy. For HuBERT and Wav2Vec2, OMP delays the collapse even more (from $\sim$270--450 to $\sim$750 rows for FSD50K).

\section{Iterative Build Experiment}
\label{sec:build}

\subsection{Design}

The iterative build experiment starts from an empty dataset and adds one row at a time. At each step, the LOO evaluation is computed on the current subset. This measures how many data instances are needed before compositionality becomes detectable.

Three build orderings are compared:

\begin{itemize}[nosep]
    \item \textbf{OMP (oracle):} adds the unselected row with the largest reconstruction residual. Uses both $S$ and $R$.
    \item \textbf{Entropy (heuristic):} adds rows from largest to smallest entropy contribution. Uses only $S$.
    \item \textbf{Random (baseline):} adds rows in a fixed random order.
\end{itemize}

\textbf{Important note on build orderings:} Both OMP and entropy build orderings require knowledge from the full dataset to decide which row to add next. OMP needs both the full $S$ and $R$ to compute reconstruction residuals for all unselected rows. Entropy needs the full $S$ to rank rows by entropy contribution. Only random build simulates genuine incremental data collection without foreknowledge.

Therefore, OMP and entropy build experiments answer: ``given access to the full dataset, what is the minimum subset needed for good reconstruction?'' They do not simulate real data collection. Random build answers: ``if we collect samples one-by-one without knowing what they will be, how much data do we need?''

\subsection{Three thresholds identified}

Analogous to removal, we identify three thresholds:

\begin{enumerate}[nosep]
    \item \textbf{values\_start:} the number of rows at which metric values first become numerically meaningful.
    \item \textbf{turning\_point:} the row count where cosine similarity reaches its minimum before improving.
    \item \textbf{gap\_positive\_from:} the row count from which the compositionality gap becomes positive.
\end{enumerate}

These form a natural ordering: values\_start $<$ turning\_point $<$ gap\_positive\_from.

\subsection{Results}

\subsubsection{OMP build: values\_start and turning\_point}

OMP build thresholds are model-dependent because OMP uses both $S$ and $R$. CLAP OMP build produces monotonic cosine similarity in all five configs (no turning point), and values become meaningful very early (4--70 rows depending on config size).

\begin{table}[h]
\centering
\caption{OMP build: values\_start and turning\_point (rows) by model.}
\label{tab:build_omp_vstp}
\begin{tabular}{l rr rr rr}
\toprule
 & \multicolumn{2}{c}{CLAP} & \multicolumn{2}{c}{HuBERT} & \multicolumn{2}{c}{Wav2Vec2} \\
\cmidrule(lr){2-3} \cmidrule(lr){4-5} \cmidrule(lr){6-7}
Config & start & t.p. & start & t.p. & start & t.p. \\
\midrule
FSD50K       & 70   & None & 64   & 59  & 21   & 52  \\
FSD50K Env   & 28   & None & 10   & 27  & 45   & 25--30 \\
FSD50K Music & 13   & None & 15   & 20  & 10   & 21  \\
FSD50K Speech& 18   & None & 18   & 16  & 22   & 19  \\
CHiME-Home   & 4    & None & 6    & 5   & 6    & 11  \\
\bottomrule
\end{tabular}
\end{table}

\subsubsection{Entropy build: values\_start and turning\_point}

Entropy build uses only $S$ to determine row ordering, so values\_start is model-independent within each config. The turning point, however, depends on cosine similarity in representation space and therefore varies across models.

\begin{table}[h]
\centering
\caption{Entropy build: values\_start and turning\_point (rows) by model. The values\_start column is model-independent.}
\label{tab:build_entropy_vstp}
\begin{tabular}{l r rrr}
\toprule
 & & \multicolumn{3}{c}{turning\_point} \\
\cmidrule(lr){3-5}
Config & values\_start & CLAP & HuBERT & Wav2Vec2 \\
\midrule
FSD50K       & 111 & 109  & 109  & 109 \\
FSD50K Env   & 53  & 64   & 62   & 60  \\
FSD50K Music & 22  & 23   & 27   & 27  \\
FSD50K Speech& 22--23 & 29 & 25   & 25  \\
CHiME-Home   & 6   & 10   & None & None \\
\bottomrule
\end{tabular}
\end{table}

For CHiME-Home, HuBERT and Wav2Vec2 entropy build produce monotonic cosine similarity (no turning point), while CLAP has a turning point at 10 rows.

\subsubsection{Random build: values\_start and turning\_point}

Random build also uses a model-independent row ordering, so values\_start is the same for all models.

\begin{table}[h]
\centering
\caption{Random build: values\_start and turning\_point (rows) by model. The values\_start column is model-independent.}
\label{tab:build_random_vstp}
\begin{tabular}{l r rrr}
\toprule
 & & \multicolumn{3}{c}{turning\_point} \\
\cmidrule(lr){3-5}
Config & values\_start & CLAP & HuBERT & Wav2Vec2 \\
\midrule
FSD50K       & 95  & 130  & 129  & 130 \\
FSD50K Env   & 45  & 67   & 85   & 85  \\
FSD50K Music & 16  & 18   & 18   & 18  \\
FSD50K Speech& 15  & 12   & 28   & 28  \\
CHiME-Home   & 6   & 8    & 8    & 8   \\
\bottomrule
\end{tabular}
\end{table}

\subsubsection{Gap-positive threshold is model-dependent even for model-independent orderings}

Even though entropy and random orderings select the same rows for all models, the gap\_positive\_from threshold differs because it depends on the representation space:

\begin{table}[h]
\centering
\caption{Entropy build: gap\_positive\_from (rows) by model.}
\label{tab:build_gap_entropy}
\begin{tabular}{lrrr}
\toprule
Config & CLAP & HuBERT & Wav2Vec2 \\
\midrule
FSD50K       & 121  & 250  & 350 \\
FSD50K Env   & 64   & 175  & 175 \\
FSD50K Music & 24   & 55   & 69 \\
FSD50K Speech& 31   & 43   & 41 \\
CHiME-Home   & 20   & 38   & 46 \\
\bottomrule
\end{tabular}
\end{table}

\begin{table}[h]
\centering
\caption{Random build: gap\_positive\_from (rows) by model.}
\label{tab:build_gap_random}
\begin{tabular}{lrrr}
\toprule
Config & CLAP & HuBERT & Wav2Vec2 \\
\midrule
FSD50K       & 128  & 200  & 270 \\
FSD50K Env   & 70   & 165  & 225 \\
FSD50K Music & 32   & 76   & 80 \\
FSD50K Speech& 14   & 45   & 56 \\
CHiME-Home   & 9    & 25   & 26 \\
\bottomrule
\end{tabular}
\end{table}

CLAP achieves a positive gap with far fewer rows than HuBERT or Wav2Vec2 across all configs and all orderings. Random build generally requires fewer rows than entropy build for CLAP (e.g., FSD50K Speech: 14 vs 31 rows), but the pattern varies for HuBERT and Wav2Vec2.

\subsubsection{OMP build gap-positive is very late for non-CLAP models}

Despite OMP being data-efficient for reaching meaningful values and turning points, the gap\_positive threshold for OMP is often high for HuBERT and Wav2Vec2:

\begin{table}[h]
\centering
\caption{OMP build: gap\_positive\_from (rows) by model.}
\label{tab:build_gap_omp}
\begin{tabular}{lrrr}
\toprule
Config & CLAP & HuBERT & Wav2Vec2 \\
\midrule
FSD50K       & 117  & 725   & 775 \\
FSD50K Env   & 75   & 400   & 400 \\
FSD50K Music & 43   & 75    & 78 \\
FSD50K Speech& 38   & 63    & 58 \\
CHiME-Home   & 20   & 35    & 39 \\
\bottomrule
\end{tabular}
\end{table}

For FSD50K and FSD50K Environmental, HuBERT and Wav2Vec2 OMP need more than half the full dataset to achieve a positive gap. This is because OMP selects rows that maximise reconstruction accuracy for the specific model. For HuBERT and Wav2Vec2, whose representations do not compose as additively, the OMP-selected subset achieves good reconstruction for the real pairing but also for permuted baselines, making the gap hard to establish.

\section{By-Category CCA and LOO Analysis (FSD50K)}
\label{sec:category}

In addition to the full FSD50K evaluation, we run CCA and LOO analysis separately for each FSD50K sound category (Environmental, Music, Speech). These by-category analyses serve two purposes: (1) they test whether compositionality varies across sound domains, and (2) they provide the foundation for the category-specific iterative experiments described above.

The by-category CCA results confirm that linear alignment between representations and source labels holds within each category for all three models. The by-category LOO results show the same pattern as the full dataset: CLAP outperforms HuBERT and Wav2Vec2, especially on cosine similarity and CKA.

These by-category analyses are documented in the Jupyter notebooks:
\begin{itemize}[nosep]
    \item \texttt{FSD50K/fsd50k\_clap\_category\_analysis.ipynb}
    \item \texttt{FSD50K/fsd50k\_hubert\_category\_analysis.ipynb}
    \item \texttt{FSD50K/fsd50k\_wav2vec2\_category\_analysis.ipynb}
\end{itemize}

\section{Cross-Configuration Patterns}
\label{sec:patterns}

Several patterns hold consistently across all five configurations:

\subsection{Pattern 1: CLAP OMP monotonicity is universal}

CLAP OMP always produces monotonic cosine similarity in both removal and build. No other model-method combination shows this. This is the strongest evidence that CLAP's additive compositional structure is qualitatively different from HuBERT's and Wav2Vec2's.

\subsection{Pattern 2: Gap-positive hierarchy is CLAP, then HuBERT, then Wav2Vec2}

In every config and every ordering, CLAP achieves a positive gap with the fewest rows (build) or maintains a positive gap to the fewest rows remaining (removal). Wav2Vec2 consistently requires the most rows.

\subsection{Pattern 3: Turning points scale with effective rank}

The entropy/effcols turning point (rows remaining) is approximately equal to $\text{eff\_rank}(S)$:

\begin{table}[h]
\centering
\caption{Entropy removal turning point (cosine similarity minimum, rows remaining) per model.}
\label{tab:tp_by_model}
\begin{tabular}{lrrr}
\toprule
Config & CLAP & HuBERT & Wav2Vec2 \\
\midrule
FSD50K       & 140 & 140 & 140 \\
FSD50K Env   & 92  & 92  & 90  \\
FSD50K Music & 34  & 30  & 30  \\
FSD50K Speech& 23  & 26  & 28  \\
CHiME-Home   & 6   & 8   & 7   \\
\bottomrule
\end{tabular}
\end{table}

\begin{table}[h]
\centering
\caption{Turning point vs effective rank. The turning point is the median across models from Table~\ref{tab:tp_by_model}.}
\label{tab:tp_vs_effrank}
\begin{tabular}{lrrr}
\toprule
Config & eff\_rank($S$) & Median TP (rows) & Ratio \\
\midrule
FSD50K       & $\sim$150 & 140  & $\sim$0.9 \\
FSD50K Env   & $\sim$97  & 92   & $\sim$0.9 \\
FSD50K Music & $\sim$25  & 30   & $\sim$1.2 \\
FSD50K Speech& $\sim$21  & 26   & $\sim$1.2 \\
CHiME-Home   & $\sim$7   & 7    & $\sim$1.0 \\
\bottomrule
\end{tabular}
\end{table}

The ratio is consistently in the range 0.9--1.2, confirming that the turning point occurs when $q \approx \text{eff\_rank}(S)$. Note that CLAP's turning point can differ substantially from HuBERT and Wav2Vec2 (e.g., CHiME-Home CLAP = 3 vs HuBERT = 8), reflecting CLAP's stronger compositionality which delays the onset of rank deficiency.

\subsection{Pattern 4: Entropy and effcols are always equivalent}

In every config, entropy and effcols orderings produce identical or near-identical turning points and values\_diminish thresholds. Ordering by column coverage does not preserve rank structure better than ordering by entropy.

\subsection{Pattern 5: Small datasets show qualitatively similar behavior}

FSD50K Music (81 rows), FSD50K Speech (89 rows), and CHiME-Home (67 rows) show the same patterns as FSD50K (1289 rows), with more noise due to fewer iterations. The three-threshold structure is still identifiable.

\section{Validity of the Main Paper Results}
\label{sec:validity}

The main paper reports CCA and LOO results on the full FSD50K (1289 clips, $\mathcal{R} \approx 8.6$) and CHiME-Home (67 clips, $\mathcal{R} \approx 9.8$). Both are safely in the overdetermined regime. The iterative experiments confirm that:

\begin{itemize}[nosep]
    \item For FSD50K, the evaluation remains valid (positive gap for CLAP) down to approximately 141 rows with entropy ordering, and down to 39 rows with OMP ordering. The full dataset (1289 rows) is far from the validity boundary.
    \item For CHiME-Home, the evaluation remains valid down to approximately 6--8 rows with entropy ordering, and down to 3 rows with OMP for CLAP. The full dataset (67 rows) is safe.
    \item The same conclusions hold for the three FSD50K category subsets.
\end{itemize}

These findings establish that the main paper's claims about compositionality differences between CLAP and speech-only models are not artifacts of insufficient data.

\section{Limitations}
\label{sec:limitations}

\begin{enumerate}[nosep]
    \item \textbf{Effective rank is one of several possible measures.} Stable rank or numerical rank might give slightly different $\mathcal{R}$ values at the transition. The $\mathcal{R} \approx 1.0$--1.6 range should be understood as approximate.

    \item \textbf{Ridge regularization interacts with the transition.} The parameter $\lambda$ provides some buffer against ill-conditioning. Different $\lambda$ values would shift the boundary slightly.

    \item \textbf{OMP is an oracle, not a practical strategy.} OMP uses $R$ to select rows, but $R$ is the quantity being evaluated. OMP results provide an upper bound on data efficiency. Similarly, entropy build uses the full $S$, which may not be available in genuine incremental data collection.

    \item \textbf{The analysis is specific to the linear framework.} Non-linear evaluation methods might have different validity boundaries.

    \item \textbf{Small datasets have limited resolution.} With 67--89 rows, turning points are identified from fewer iterations and are inherently noisier.

    \item \textbf{FSD50K subsets are not independent.} The three category subsets share audio clips and labels with the full FSD50K. CHiME-Home is the independent external validation.
\end{enumerate}

\section{Summary of All Turning Points}
\label{sec:summary}

Complete turning point tables for all 5 configurations $\times$ 18 experiments (3 models $\times$ 3 orderings $\times$ 2 directions) are available in the file \texttt{all\_manual\_turning\_points.md}.

The cause analysis notebooks for each configuration are at:
\begin{itemize}[nosep]
    \item \texttt{FSD50K/Turning\_Point\_Analysis/turning\_point\_cause\_analysis.ipynb}
    \item \texttt{FSD50K\_Environmental/Turning\_Point\_Analysis/turning\_point\_cause\_analysis.ipynb}
    \item \texttt{FSD50K\_Music/Turning\_Point\_Analysis/turning\_point\_cause\_analysis.ipynb}
    \item \texttt{FSD50K\_Speech/Turning\_Point\_Analysis/turning\_point\_cause\_analysis.ipynb}
    \item \texttt{chime\_home/Turning\_Point\_Analysis/turning\_point\_cause\_analysis.ipynb}
\end{itemize}